\documentclass[aps, twocolumn, amsmath, amssymb]{revtex4}

\usepackage{color}
\usepackage{graphics} 
\usepackage{graphicx} 
\usepackage{subfigure}
\usepackage{epstopdf}
\usepackage{ulem}

\newcommand{\llabel}[1]{\label{#1}} 

\newcommand{\BQIC}{Berkeley Center for Quantum Information and Computation, Berkeley, California 94720 USA}
\newcommand{\DeptPhys}{Department of Physics, University of California, Berkeley, California 94720 USA}
\newcommand{\DeptChem}{Department of Chemistry, University of California, Berkeley, California 94720 USA}

\newcommand{\erf}[1]{Eq.~(\ref{#1})} 

\def\>{\rangle}
\def\<{\langle}
\def\t{\dagger}
\def\->{\rightarrow}
\def\=>{\implies}
\def\6{\partial}

\def\({\left(}
\def\){\right)}
\def\[{\left[}
\def\]{\right]}
\def\Tr{\text{Tr}}

\newcommand{\cut}[1]{} 

\begin{document}

\title{Single-shot deterministic entanglement between non-interacting systems with linear optics} 

\author {Leigh S. Martin$^{1,2}$}
\email {Leigh@Berkeley.edu}
\author{K. Birgitta Whaley$^{1,3}$}

\affiliation{$^1$\BQIC}
\affiliation{$^2$\DeptPhys}
\affiliation{$^3$\DeptChem}

\date{\today}

\begin{abstract} 
Measurement-based heralded entanglement schemes 
have served as the primary link between physically separated qubits in most quantum information platforms.
However, the impossibility of performing a deterministic Bell measurement with linear optics bounds the success rate of the standard protocols to at most 50\%, which means that the entanglement of the unheralded state is zero. Here we show that the ability to perform feedback during the measurement process enables unit success probability in a single shot. Our primary feedback protocol, based on photon counting retains the same robustness as the standard Barrett-Kok scheme, while doubling the success probability even in the presence of loss. In superconducting circuits, for which homodyne detectors are more readily available than photon counters, we give another protocol that can deterministically entangle remote qubits given existing parameters. In constructing the latter protocol, we derive a general expression for locally optimal control that applies to any continuous, measurement-based feedback problem. 
\end{abstract}


\maketitle

Scalable proposals for quantum computation typically call for discrete modules,
which must be entangled to obtain an advantage over classical processors\cite{divincenzo2000physical,duan2004scalable, Devoret2013superconducting}. 
Heralded schemes based on single photon detection have become the method of choice for entangling non-interacting or distant quantum systems, with implementation in trapped ions, cold neutral atoms, nitrogen vacancy centers, quantum dots and superconducting circuits systems\cite{moehring2007entanglement, Hofmann2012, Bernien2013, slodivcka2013atom, narla2016robust, delteil2016generation,stockill2017phase}. Such methods are also promising for interfacing different types of matter qubits in hybrid quantum information processing, since otherwise incompatible qubits can often be made to emit indistinguishable photons through frequency tuning\cite{Bernien2013} or frequency conversion\cite{tchebotareva2019entanglement}. Unfortunately, due to the impossibility of performing a deterministic Bell measurement in this setting, the success rate is intrinsically limited to 50\%\cite{Calsamiglia2001maximum}. While this limitation is small compared to the effect of loss for truly remote (\textit{i.e.}, kilometer-separated) systems, it becomes significant in the intermediate-range scales encountered in modular computing platforms, where loss may be greatly reduced\cite{Roch2014}.

In this work, we present two methods for generating \textit{deterministic} entanglement using two commonly-used measurement schemes. The first, based on photon counting, retains the same high degree of robustness to loss and other imperfections afforded by the standard 
method of Barrett and Kok\cite{barrett2005efficient}. 
We show that this approach is also moderately robust to feedback delay, which reduces the success probability without affecting the fidelity of state preparation. The second scheme replaces photon counters with standard homodyne detection. This detection scheme is well suited to superconducting circuits, where photon counters with temporal resolution have yet to be demonstrated. While the resulting protocol is less robust to loss, we show that it can 
nevertheless create entanglement deterministically under currently achievable parameters. 


In the following, we first describe our protocol based on photon counters, and show that it achieves unit success probability in the ideal case. We then evaluate its performance in the presence of loss, loss asymmetry, path length fluctuations, feedback delay and non-Markovian effects arising from use of Purcell enhancement cavities.  
Given typical experimental parameter regimes, none of these imperfections measurably degrade the fidelity of state preparation. 
In the second half of the paper, we derive an alternative protocol based on homodyne detection, which also deterministically reaches unit fidelity under ideal conditions. An important theoretical contribution of this work comes in the analysis of this protocol in the presence of loss and feedback delay, for which we extend the locally optimal feedback scheme of Ref. \cite{Zhang2018Locally} to permit multiple non-commuting feedback Hamiltonians. This result allows one to derive the locally optimal feedback Hamiltonian for any measurement-based control problem with a simple, closed-form expression. We also extend this result to retain local optimality in the presence of delay to first order.

\begin{figure}
\centering
\includegraphics[width =0.45\textwidth]{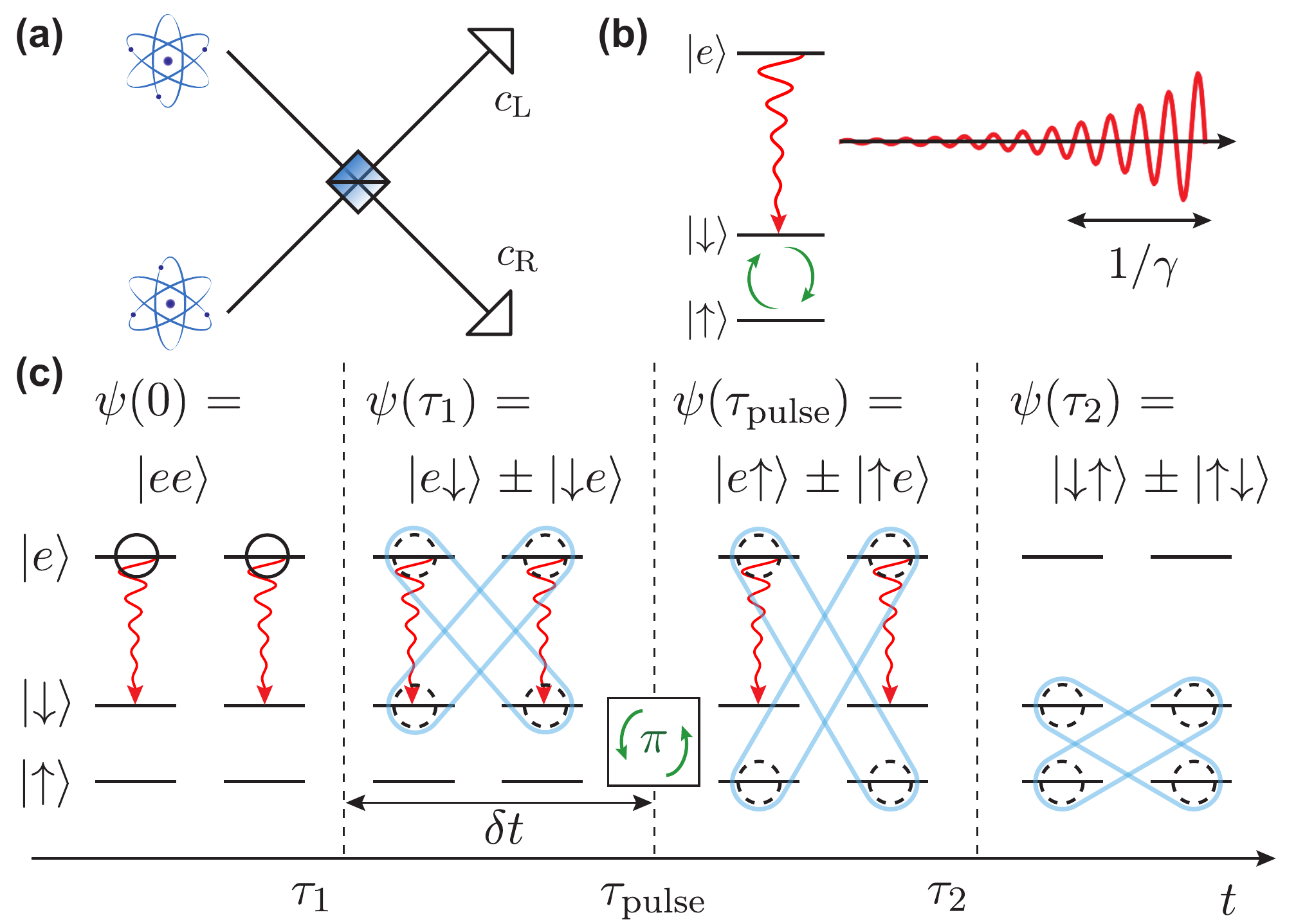}
\caption{\textbf{(a)} Experimental system for entanglement generation. The spontaneous emission from two separated atoms is interfered on a beam splitter. Two photon counters are placed at the beam splitter outputs. \textbf{(b)} Assumed level structure for the atoms. The $|e\>$ state decays to $|{\downarrow}\>$ and emits a photon of duration $\mathcal{O}(1/\gamma)$. Feedback is then applied on the $|{\downarrow}\>\leftrightarrow|{\uparrow}\>$ transition. \textbf{(c)} Robust entanglement scheme using photon counters. At $t=0$, the system is initialized in $|ee\>$, which radiatively decays to $|{\downarrow}\>$. At $t=\tau_1$, a single photon is detected, which projects the system into an entangled state. At a later time $\tau_\text{pulse}$ set by the feedback delay, $\pi$ pulses are applied to the $|{\downarrow}\>\leftrightarrow|{\uparrow}\>$ transition. Once the second photon is detected at $\tau_2$, the system collapses to a Bell state.}
\label{fig:PhotonCountingSetup}
\end{figure}

\textbf{Photon counting scheme.} We begin with the more intuitive and robust protocol based on photon counting. Fig. \ref{fig:PhotonCountingSetup}a depicts the basic experimental system. Two atoms spontaneously emit onto a beam splitter, and two photon counters detect the signal at the beam splitter's outputs. The simplest relevant atomic system consists of a pair of long-lived ground state sublevels $|{\uparrow}\>$ and $|{\downarrow}\>$, and an additional optically active excited state $|e\>$ that decays to $|{\downarrow}\>$ (Fig. \ref{fig:PhotonCountingSetup}b). 
If the emitted photons are indistinguishable, then the atom-photon entanglement inherent to the spontaneous emission process can be converted to atom-atom entanglement via entanglement swapping\cite{coecke2004logic}. This process relies on the beam splitter erasing information about which photon came from which atom. 
As discussed in the introduction, existing schemes are limited to a success probability of 50\% \cite{Calsamiglia2001maximum}. For example, in the standard Barrett-Kok protocol, in which one initially prepares $(|e\> + |{\uparrow}\>) \otimes (|e\> + |{\uparrow}\>)$ (ignoring normalization), detection of zero or two photons projects the system into the separable states $|{\uparrow \uparrow}\>$ or $|{\downarrow \downarrow}\>$ respectively. Only detection of exactly one photon projects the system into the entangled state $|{\uparrow \downarrow}\> \pm |{\downarrow \uparrow}\>$.

Our protocol for deterministic entanglement generation is depicted in Fig. \ref{fig:PhotonCountingSetup}c. We take the spontaneous emission rates $\gamma$ to be equal. The atoms are initialized in $|ee\>$, and the feedback controller waits for a single photon. During this time interval, the atoms evolve trivially under the non-Hermitian Hamiltonian $H = -i \sum_i \gamma \sigma_i ^\t \sigma_i/2$, where $\sigma = |{\downarrow}\>\<e|$. Both atoms remain in $|e\>$ until detection of a single photon at one of the detectors. Since the photons have undergone a beam splitter interaction, the atomic jump operators associated with the left and right photon counters are coherent combinations of the atomic ladder operators\cite{barrett2005efficient}
\begin{align}
	c_\text{L/R} = \frac{\sigma_1 \pm \sigma_2}{\sqrt{2}}.
\end{align}
The state after the first detection event is an entangled state, $|\psi(\tau_1)\> = |e{\downarrow}\> \pm |{\downarrow}e\>$.
In the absence of feedback, a subsequent photon detection occurring a time $\mathcal{O}(1/\gamma)$ later destroys this entanglement, leaving the system in $|{\downarrow \downarrow}\>$. Note that the second photon arrives at the same detector as the first, due to Hong-Ou-Mandel interference \cite{hong1987measurement}. We can prevent loss of entanglement by applying a $\pi$ pulse on the $|{\uparrow}\>\leftrightarrow |{\downarrow}\>$ transition before detection of the second photon, which results in the final state $|{\downarrow\uparrow}\> \pm |{\uparrow \downarrow}\>$. This pulse also destroys Hong-Ou-Mandel interference, so that the second photon has equal probability to arrive at either detector. Regardless of the measurement outcome, the scheme prepares a unit-fidelity Bell state. 

The above protocol is remarkably robust to commonly encountered experimental imperfections. The primary limit to remote entanglement generation is loss. Fortunately, as long as the dark count rate of the photon counters is low, detection of two photons guarantees collapse into the target state with fidelity that is independent of loss. The success rate is therefore $(1-P_\text{loss})^2$, 
a significant improvement over the value of $(1-P_\text{loss})^2/2$ for the Barrett-Kok scheme. The final state fidelity is also unaffected by asymmetry in the loss along the two paths leading to the beam splitter, as well as path length fluctuations, so long as these parameters change slowly compared to $1/\gamma$\cite{SupplementaryMaterials}.
Physically, these asymmetries cancel in the final state because we detect exactly one photon from each atom.

The necessity of fast feedback operations requires several novel considerations. Feedback delay presents the most readily apparent challenge. In particular, the protocol will fail if the second photon detection occurs before the feedback pulse is completed.
Since the photon wave packet decays exponentially, the probability of success is therefore $P_\text{success} = \gamma \int_{\delta t}^\infty e^{-\gamma \delta t}dt = e^{-\gamma \delta t}$, where $\delta t$ is the delay time.
As long as $\delta t$ includes the pulse duration, feedback delay does not decrease the fidelity of the final state. Another more technical detail is that the increased detection bandwidth required for high temporal resolution potentially leads to 
a greater number of dark counts from thermal photons. This issue merits consideration on a system-by-system basis, since thermal background and dark count rate depend on specifics of the photon counters and energy of the spontaneously emitted light.


The final imperfection that we consider is the effect of putting the atoms in cavities, which can increase the photon collection efficiency and enhance the spontaneous emission rate via the Purcell effect\cite{van2018optimal}. In the context of quantum feedback, cavities introduce a form of non-Markovianity that resembles feedback delay. In the most commonly encountered limit of $\gamma \ll \kappa$, the final state is approximately\cite{SupplementaryMaterials}
\begin{align}
    |\psi(\infty)\> \propto \frac{|{\uparrow \downarrow}\> + |{\downarrow \uparrow}\>}{\sqrt{2}} + \sqrt{2} e^{-\kappa(\tau_2-\tau_\text{pulse})/2} |{\uparrow \uparrow}\>
\end{align}
if the same photon counter registers both photons, and $|{\downarrow \uparrow}\> - |{\uparrow \downarrow}\>$ otherwise. Here $\kappa$ is the cavity bandwidth.
Intuitively, the population in $|{\uparrow \uparrow}\>$ arises when both atomic excitations have swapped into the cavity at the moment when the feedback pulse is applied. Note that since $\tau_2-\tau_\text{pulse} = \mathcal{O}(1/\gamma)$, this correction term is typically exponentially small in $\kappa/\gamma$.

\begin{figure}
\centering
\includegraphics[width =0.45\textwidth]{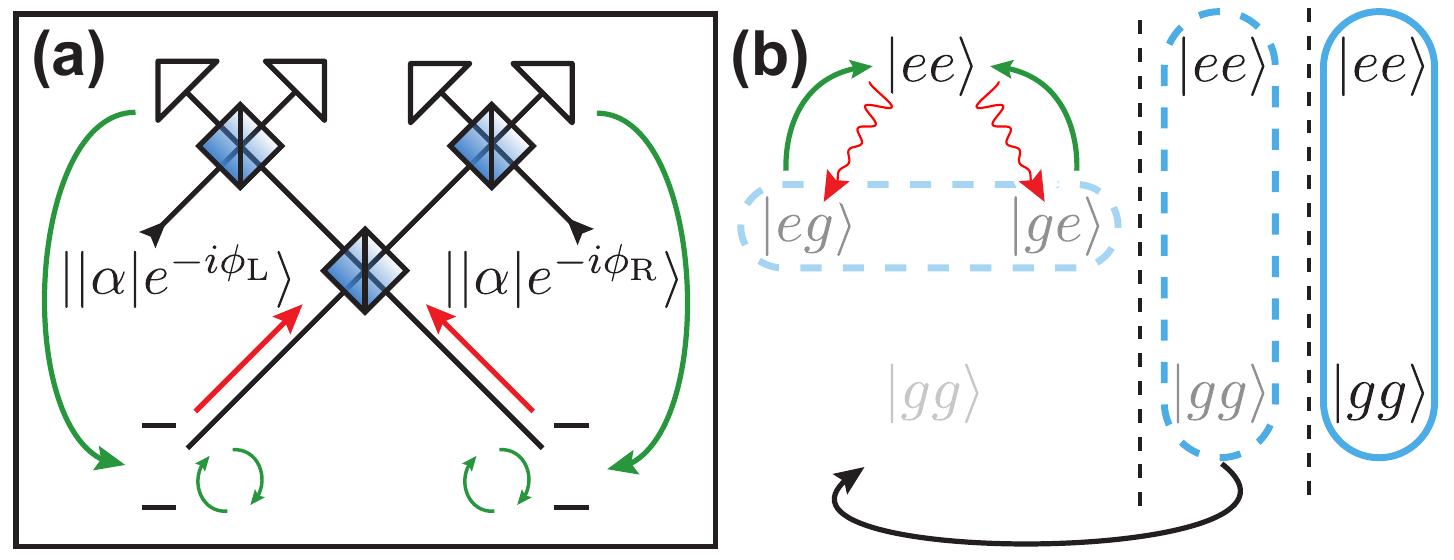}
\caption{\textbf{(a)} Physical system for the homodyne detector version of our protocol. The system is that of Fig. \ref{fig:Homodyne-Setup}a except with the photon counters replaced by homodyne detectors or phase-sensitive parametric amplifiers. Detectors measure field quadratures parameterized by $\phi_i$. Homodyne detection is depicted, implemented by feeding strong coherent states $|\alpha\>$ onto beam splitters. The atomic level structure is now that of a qubit, with excited and ground states $|e\>$ and $|g\>$. \textbf{(b)} Intuitive description of the homodyne protocol. Atoms begin in $|ee\>$. Any fluorescence detection leads to entanglement, so long as the homodyne detector phases are $\pi/2$ out of phases ($|\phi_\text{L} - \phi_\text{R}| = \pi/2$). Feedback maintains the state in Schmidt form, maximizing the population in $|ee\>$. The final Bell state $|ee\> + |gg\>$ is reached deterministically, independent of the measurement record.}
\label{fig:Homodyne-Setup}
\end{figure}

\textbf{Homodyne detection scheme.} Photon counters are a significant challenge for some systems, including superconducting circuits.
However superconducting phase preserving amplifiers are readily available with quantum efficiencies over 50\%\cite{eddins2018transamp,walter2017rapid}, opening the possibility for feedback via continuous quadrature detection\cite{campagne2016observing}.
The requisite Hong-Ou-Mandel interference observed via quadrature detection has already been demonstrated in superconducting circuits that were too far apart to interact directly\cite{lang2013probing}. 
A recent theoretical study 
considered entanglement formation in this context, and confirmed the upper bound success rate of 50\%
(quantified as average concurrence)\cite{lewalle2019entanglement, lewalle2019diffusive}.
We now present a deterministic protocol analogous to that of Fig. \ref{fig:PhotonCountingSetup}a 
but using parametric amplifiers, or equivalently homodyne detectors, instead of photon counters. The corresponding experimental design is shown in Fig. \ref{fig:Homodyne-Setup}a.

We begin with the stochastic master equation for homodyne detection\cite{JacobsSteck2006}. This protocol only requires two-level atoms, so for simplicity we work with only two states, $|e\>$ and $|g\>$. An equation of motion analogous to that generated by the non-Hermitian Hamiltonian $H$ from the photon-counting case describes the evolution under continuous homodyne measurement
\begin{align} \label{eq:SSEHomodyne}
d|\psi\> = \Big[\sum_{i=\{\text{L,R}\}} &-\frac{\gamma}{2} (c_i^\t c_i - 2 \<x_i\>c_i + \<x_i\>^2)dt \\ \nonumber
&+ \sqrt{\gamma}(c_i-\< x_i \>)dW_i \Big] |\psi\>,
\end{align}
where $c_\text{L/R} = \exp(-i\phi_\text{L/R})(\sigma_1 \pm \sigma_2)/\sqrt{2}$, with $\sigma = |g\>\<e|$. Here $x_i = (c_i+c_i^\t)/2$ are the measurement axes, and $\phi_\text{L/R}$ are the local oscillator phases for the left and right detectors. \erf{eq:SSEHomodyne} contains $H$ as its first term, but also contains nonlinear terms $\<x_i\>|\psi\>$ which preserve normalization, as well as stochastic terms proportional to $dW_i$ that are analogous to quantum jumps but that instead generate diffusive motion. Each $dW_i$ is a zero mean Gaussian random variable with variance $dt$, and may be determined from the associated measurement records $dr_i = 2\sqrt{\gamma} \<x_i\>dt + dW_i$.

To derive a feedback protocol for entanglement generation, we derive an equation of motion for the concurrence\cite{martin2017optimal, Wootters1998}, an entanglement measure with a simple analytic expression in the case of two qubits. We also show that it is 
locally 
optimal to apply feedback rotations on each qubit to maintain the state in Schmidt form up to a phase at all times \textit{i.e.}, $|\psi(t)\> = e^{i\phi_\lambda}\lambda_e(t) |ee\> + \lambda_g(t) |gg\>$\cite{Nielsen2010}. 
We illustrate the physical intuition behind this criteria in Fig. \ref{fig:Homodyne-Setup}b. Among all initially separable states, $|ee\>$ offers the largest spontaneous emission rate. After a short time interval, the atoms partially decay into $|eg\>$ and $|ge\>$, which can generate a small amount of entanglement. Rotating this state back to Schmidt form again maximizes the population in $|ee\>$, which then maximizes the rate of entanglement generation at the next time step. Continuously repeating this procedure will eventually yield the maximally entangled state $e^{i\phi_\lambda}|ee\> + |gg\>$.

\begin{figure}
\centering
\includegraphics[width =0.48\textwidth]{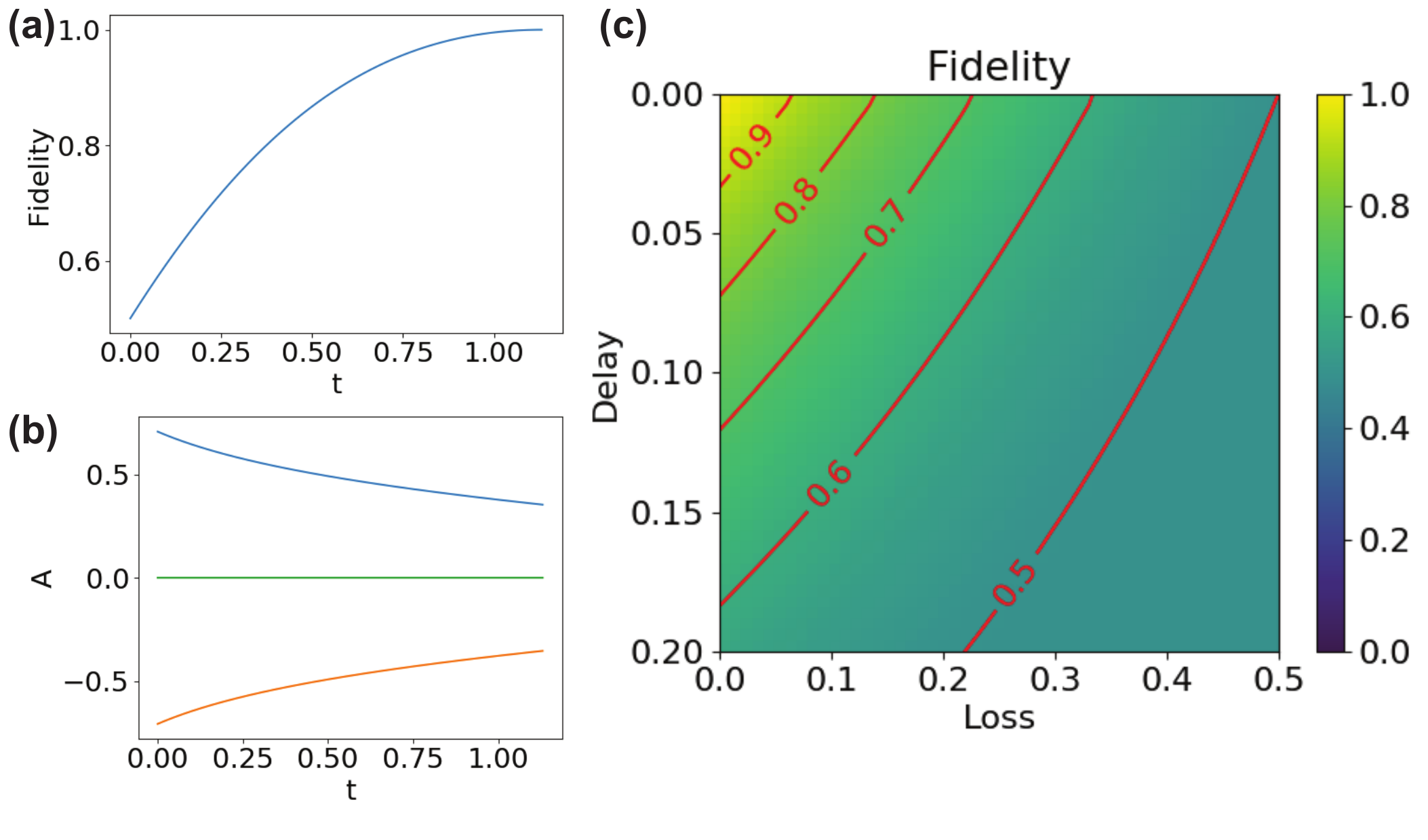}
\caption{\textbf{(a)} Fidelity as a function of time for the homodyne protocol for the ideal case $\eta = 1$ and $\delta t = 0$. The fidelity hits exactly 1 in finite time. \textbf{(b)} Feedback coefficients $A_{ij}$ for the ideal case. The top curve (blue) contains $A_{\text{L}, \sigma_{x,1}}$, $A_{\text{L}, \sigma_{x,2}}$ and $A_{\text{R}, \sigma{y,2}}$, while the bottom curve (orange) contains $A_{\text{R}, \sigma_{y,1}}$. All other matrix elements are zero (green). \textbf{(c)} Fidelity as a function of loss and feedback delay. Contours show lines of constant fidelity. No entanglement is generated below the $0.5$ contour.}
\label{fig:HomodyneFidelities}
\end{figure}

For a state in Schmidt form, the equation of motion for concurrence is quite simple. 
Choosing $\lambda_e \geq \lambda_g$, the result for $\mathcal{C} \neq 0$ is \cite{SupplementaryMaterials}
\begin{align}   \label{eq:HOMFdC}
&\frac{d\mathcal{C}}{dt} 
= 
\sin(\phi_\text{L} - \phi_\text{R})\sin(\phi_\text{L} + \phi_\text{R} - \phi_\lambda)(\sqrt{1-\mathcal{C}^2} + 1) - \mathcal{C}
\end{align}
and for $\mathcal{C}=0$ is $d\mathcal{C}/dt =  2|\sin(\phi_\text{L} - \phi_\text{R})|$.
We have set $\gamma = 1$ for simplicity of notation. It is evident that the amount of entanglement generated depends strongly on the phase of the local oscillators $\phi_\text{L/R}$, as has also been observed in the absence of feedback\cite{lewalle2019entanglement, lewalle2019diffusive}. It is interesting to note that the concurrence evolves deterministically despite the randomness of the measurement process. This phenomenon has been observed in a number of other feedback protocols\cite{Jacobs2003,HPFPRA, martin2017optimal}. 
For the optimal choice of $|\phi_L-\phi_R| = \phi_\text{L} + \phi_\text{R} - \phi_\lambda = \pi/2$, the system reaches exactly unit concurrence at $t = \pi/4 + \ln(2)/2 \approx 1.13$. The fidelity as a function of time is plotted in Fig. \ref{fig:HomodyneFidelities}a.
 
Unlike in the photon counting protocol above, homodyne detectors do not offer a clear signal that heralds the arrival of two photons. Consequently, the protocol is more sensitive to loss. To assess the effect of loss, we extend the PaQS feedback equations of Ref. \cite{Zhang2018Locally} to solve for the locally optimal protocol in the presence of imperfections.
We also derive a correction to these equations which models feedback delay, valid to first order in $\delta t \gamma$. 

A significant result of this paper 
is the extension of the PaQS feedback equations to handle multiple non-commuting feedback Hamiltonians, a necessary capability for the present system. 
We consider continuous measurement of a set of operators $M_i$ with efficiencies $\eta_i$ and associated measurement records $dr_i(t) = \sqrt{\eta_i} \<M_i + M_i^\t\> dt + dW_i$. Given the ability to apply a set of feedback Hamiltonians $H_j$ with strengths given by $\sum_i A_{ij} dW_i(t)$, the feedback master equation obtained by averaging over the measurement record is\cite{Wiseman2009book, SupplementaryMaterials}
\begin{align} \label{eq:FeedbackSME}
	\frac{d\rho}{dt} &= - i \sum_j B_j [H_j, \rho] \\ \nonumber
	&~~~ + \sum_i \left[ \mathcal{D}[M_i] \rho 
	- i \sqrt{\eta_i} [\tilde{H}_i, \mathcal{H}[M_i]\rho] + \mathcal{D}[\tilde{H}_i]\rho \right], \\ \nonumber
	\tilde{H}_i &\equiv \sum_j A_{ij} H_j,
\end{align}
where $\mathcal{D}[X]\rho \equiv X\rho X^\t - [X^\t X \rho + \rho X^\t X]/2$ is the standard Lindblad dissipator, $\mathcal{H}[X]\rho \equiv X\rho + \rho X^\t - \<X + X^\t\> \rho$,
and we also apply each $H_j$ with an amplitude $B_j$ that is independent of the measurement record prior to time $t$.
The optimal values of the feedback coefficients $A_{ij}$ and $B_j$ can be determined by maximizing the expectation value with respect to some desired target operator $X_T$, although one requires higher-order terms beyond those in \erf{eq:FeedbackSME} (see \cite{SupplementaryMaterials} for details)
\begin{align}   \label{eq:PaQSCoefs}
A &= i a c^{-1}, ~~~ B = i b c^{-1}, \textrm{where} \\ \nonumber
a_{ij} &= \sqrt{\eta_i} \text{Tr}\left[ X_T [H_j, \mathcal{H}[M_i]\rho] \right] \\ \nonumber
b_j &= \\ \nonumber
&\sum_i \text{Tr}\left[ X_T [H_j, \mathcal{D}[M_i] \rho -i \sqrt{\eta_i} [\tilde{H}_i, \mathcal{H}[M_i]\rho] + \mathcal{D}[\tilde{H}_i]\rho ] \right] \\ \nonumber
c_{ij} &= -\text{Tr}\left[X_T [H_j, [H_i, \rho]]\right].
\end{align}
This result requires that the $H_j$ form a Lie algebra, so that the vector space spanned by $\{ H_j \}$ is closed under commutation. \erf{eq:PaQSCoefs} allows one to compute the locally optimal feedback Hamiltonian for any measurement-based control problem by using a complete matrix basis for $H_j$.

To generate a Bell state, we set the target operator $X_T$ to be a projection operator onto the target state $|ee\> + |gg\>$ and the measurement operators as $M_i = \{c_\text{L}, c_\text{R}\}$. We restrict to local feedback rotations by using $H_j = \{\sigma_{x,1}, \sigma_{y,1}, \sigma_{z,1}, \sigma_{x,2}, \sigma_{y,2}, \sigma_{z,2} \}$, and model loss by setting $\eta_1 = \eta_2 = \eta_\text{loss}$. We also include the above-mentioned correction for feedback delay. 
Fig. \ref{fig:HomodyneFidelities}c shows the fidelity as a function of loss and delay, restricted to $0.5 < \eta_\text{loss} \leq 1$. We also show the feedback coefficients $A_{ij}$ and $B_j$ as a function of time in Fig. \ref{fig:HomodyneFidelities}b for the ideal case $\eta_\text{loss}=1$ and $\delta t=0$. The optimal solutions for $A$ are such that the effective feedback Hamiltonians are $\sigma_{x,1} + \sigma_{x,2}$ and $\sigma_{y,2} - \sigma_{y,1}$, applied proportional to the $c_\text{L}$ and $c_\text{R}$ measurement records, respectively. 
For zero feedback delay ($\delta t = 0$), the optimal values for $B$ are
zero, while for finite feedback delay the values of $B_j$ are non-zero only for the $\sigma_x$ and $\sigma_y$ rotations. 


Since \erf{eq:FeedbackSME} averages over the measurement outcome at each time step, the simulated protocol is only locally optimal for the average trajectory\cite{HPFPRA}. It is possible that higher fidelities may be achievable if one applies feedback conditioned on the entire measurement record, which would result in a non-Markovian feedback protocol\cite{Zhang2018Locally} (note that \erf{eq:PaQSCoefs} remains valid in this case). 
Using Fig. \ref{fig:HomodyneFidelities}c, we note that using the quantum efficiency of $0.66$ reported in Ref. \cite{walter2017rapid}, the maximum tolerable delay is $\sim 0.13\gamma$. Given the feedback delay of 374 ns reported in Ref. \cite{martin2019adaptive}, it should be possible to reach the threshold for entanglement (0.5 fidelity) with existing superconducting circuit technology, so long as $\gamma < 0.13/0.374 = 350$ kHz.

\textbf{Conclusion and outlook} 
We have presented two protocols for deterministic entanglement of remote qubits based on feedback applied during the measurement process. The first protocol, based on photon counting doubles the success rate of standard heralded entanglement schemes.
High state preparation fidelity is unaffected by loss, feedback delay and does not require interferometric stability between remote nodes.
The second protocol uses homodyne measurements and continuous feedback. To determine the form of the latter, we have derived a general expression for locally optimal feedback protocols permitting multiple non-commuting measurement operators. This equation allows one to explicitly calculate the best feedback Hamiltonians for virtually any given measurement-based control problem.

Although we have only considered systems involving two qubits, our methods are applicable to multipartite entanglement generation as well. In particular, it should be possible to create many different types of entangled states using photon counters and networks of beam splitters. If one initializes each atom in $|e\>$, feedback schemes based on photon counting will retain robustness to loss.

Due to the sensitivity of our protocols to feedback delay, they are mostly suitable for networking of spatially nearby modules, rather than for long-distance quantum communication. However, there are technologically important cases in which they could significantly augment even truly remote applications. For example, Erbium-doped yttrium orthosilicate (Er$^{3+}$: Y$_2$SiO$_5$) has recently garnered interest as a promising solid state defect for quantum networks due to its long coherence time and emission in the telecom band\cite{ranvcic2018coherence}. With an excited state lifetime on the order of milliseconds\cite{hiraishi2019optical}, photons emitted via optical spontaneous emission span many kilometers, permitting feedback between widely separated nodes.

In addition to the unavoidable imperfections such as loss that we have considered, others exist that will require a more system-specific analysis. For instance, whenever a thermal background is present, there is a tradeoff between photodetector bandwidth and dark count rate.
The challenge of implementing fast pulses during the decay process also depends greatly on the qubits used. Nevertheless, given the high degree of robustness and the ubiquitous nature of the experimental scenario considered in this work, we expect these two deterministic entanglement generation protocols to have applications in a wide range of quantum information platforms. 



\begin{acknowledgements}
We thank Machiel Blok, Tian Zhong and Howard Wiseman for useful conversations. The effort of LM was supported by grants from the National Science Foundation Grant No. (1106400) and the Berkeley Fellowship for Graduate Study. KBW was supported by Laboratory Directed Research and Development (LDRD) funding from Lawrence
Berkeley National Laboratory, provided by the U.S. Department of Energy, Office of Science under Contract No. DE-AC02-05CH11231.
\end{acknowledgements}

\bibliography{Bibliography}

\onecolumngrid
\section{Supplemental Materials}

\subsection{Detailed derivation of the homodyne protocol under ideal conditions}
\label{sec:SMHomodyne}

Using the methods of section \ref{sec:SMPaQS} and choosing $X_T$ to be a projection operator onto the Bell state $(|ee\> + |gg\>)/\sqrt{2}$ yields the 
homodyne protocol presented in the main text. However 
it would be desirable to understand why this particular choice is preferable among all maximally entangled two-qubit states. In fact, one could argue that $|ee\>+|gg\>$ is a somewhat unnatural target, given that the quantum jump operators $c_\text{L/R}$ generate coherence and entanglement in the $|eg\>$/$|ge\>$ subspace. 

A more natural way to solve the optimization problem is to choose an entanglement measure as our figure of merit, to avoid arbitrarily selecting out a single target state. In this section, we derive the homodyne-based entangling protocol by computing an equation of motion for the concurrence\cite{Hil.Woo-1997, Wootters1998} and finding the protocol that maximizes it. The concurrence of two qubits in a pure state is defined as
\begin{equation}
\label{eq:concurrence}
\mathcal{C} \equiv |\langle \psi^*| \sigma_y \otimes \sigma_y |\psi \rangle|
\end{equation}
Our goal is first to see how the concurrence evolves under continuous measurement, and then to determine a feedback protocol that maximizes it. A convenient state parameterization for this purpose is the Schmidt decomposition\cite{Nielsen2010, martin2017optimal}. By expressing the Schmidt coefficients in terms of the concurrence, we can write a general two-qubit state as\cite{martin2017optimal}
\begin{align}
\label{eq:StateParam}
\psi &(\mathcal{C}, \theta_1, \theta_2, \phi_{\mathcal{C},1}, \phi_{\mathcal{C},2}, \theta_{z,1}, \theta_{z,2})  \\ \nonumber
&= U_1 \otimes U_2 \Bigg{[} \sqrt{\frac{1+\sqrt{1-\mathcal{C}^2}}{2}} |ee\rangle - \sqrt{\frac{1-\sqrt{1-\mathcal{C}^2}}{2}} |gg\rangle \Bigg{]} \\ \nonumber
&U_i \equiv \exp(-i \theta_{z,i} \sigma_z/2) \exp(-i \theta_i \sigma_y/2) \exp(-i \phi_{\mathcal{C},i} \sigma_z/2), \\ \nonumber
\end{align}
where we have written $U_i$ in terms of the Euler angles $\{\phi_{\mathcal{C},i}, \theta_i, \theta_{z,i}\}$. We decompose the local unitaries into symmetric and antisymmetric rotations by defining $\theta \equiv (\theta_1 + \theta_2)/2$, $\Delta \theta \equiv (\theta_1 - \theta_2)/2$, and likewise for $\phi_{\mathcal{C}}$ and $\theta_z$. The final expression does not depend on $\Delta \phi_\mathcal{C}$, because the state in brackets is invariant under antisymmetric rotations about $\sigma_z$. 

As we allow all local unitaries for our feedback operations, we can directly treat $\theta,~ \Delta \theta,~ \phi_\mathcal{C},~ \theta_z$ and $\Delta \theta_z$ as control variables. The only parameter not controlled by feedback is the figure of merit itself, so we only need an equation of motion for concurrence in order to study the efficacy of a given feedback protocol. Inserting $|\psi\> + |d\psi\>$ into \erf{eq:concurrence} and using \erf{eq:SSEHomodyne} for $|d\psi\>$ yields
\begin{align}
    d\mathcal{C} &= -\mathcal{C} dt - 2\mathcal{C}\left(\<x_\text{L}\> dW_\text{L} + \<x_\text{R}\> dW_\text{R}\right) \\ \nonumber
    &- \sin(\Delta \phi) \left[ \left( \sqrt{1-\mathcal{C}^2} \frac{u^2 + v^2}{2} + uv \right) \cos \alpha \sin 2\phi_\mathcal{C} + \left( \sqrt{1-\mathcal{C}^2}uv + \frac{u^2 + v^2}{2} \right) \sin \alpha \cos 2\phi_\mathcal{C} +\mathcal{C} \frac{u^2-v^2}{2} \sin \alpha \right] dt \\ \nonumber
    &\Delta \phi \equiv \phi_\text{L} - \phi_\text{R} \\ \nonumber
    &\alpha \equiv 2\theta_z + \phi_\text{L} + \phi_\text{R} \\ \nonumber
    &u \equiv \cos \theta \\ \nonumber
    &v \equiv \cos \Delta \theta,
\end{align}
%
for $\mathcal{C} \neq 0$ and 
\begin{align}
    d\mathcal{C} = \frac{1}{2}|\sin(\phi_\text{L} - \phi_\text{R})| (\cos \theta + \cos \Delta \theta)^2 dt
\end{align}
for $\mathcal{C} = 0$. We have used Ito's rule to drop terms of order higher than $dt$ and set 
the spontaneous emission rate $\gamma = 1$ for simplicity. Setting $u=v=1$ recovers \erf{eq:HOMFdC} of the main text. From the above expression, we see that the only way to obtain deterministic evolution is to prepare states with $\<x_\text{L}\> = \<x_\text{R}\> = 0$. We also see that the locally optimal protocol will necessarily have $\phi_\text{L}-\phi_\text{R} = \pm \pi/2$, so that magnitude of the second $\mathcal{O}(dt)$ term is maximized. This
derivation of the optimal detector configuration is consistent with that arrived at recently for a solely measurement-based protocol in Ref. \cite{lewalle2019entanglement}. All global maxima occur for $u=v=1$, which can be verified numerically. This parameter setting corresponds to maximizing the population in $|ee\>$. In this case, the $dW$ terms drop and $d\mathcal{C}$ reduces to
\begin{align}
    d\mathcal{C} &= -\mathcal{C} dt \mp (\sqrt{1-\mathcal{C}^2} + 1)\sin(\alpha + 2\phi_\mathcal{C}) dt \\ \nonumber
    &\Delta \phi = \pm\pi/2.
\end{align}
$d\mathcal{C}$ is maximized for $\alpha + 2\phi_\mathcal{C} = -\Delta \phi = \pm\pi/2$. With these parameter settings, the state during feedback resembles a Bell state with a relative phase determined by the homodyne measurement axes
\begin{align}
    |\psi\rangle &= \sqrt{\frac{1+\sqrt{1-\mathcal{C}^2}}{2}}|ee\> \pm i e^{i(\phi_\text{L} + \phi_\text{R})} \sqrt{\frac{1-\sqrt{1-\mathcal{C}^2}}{2}} |gg\> \\ \nonumber
    & \Delta \phi = \pm \pi/2.
\end{align}

Finally, to compute the dynamics under the above feedback protocol, we solve the deterministic differential equation
\begin{align}
    \frac{d\mathcal{C}}{dt} = 1-\mathcal{C} + \sqrt{1-\mathcal{C}^2}.
\end{align}
The solution can be found exactly by integration
\begin{align}
    t = \int \frac{d\mathcal{C}}{1-\mathcal{C}+\sqrt{1-\mathcal{C}^2}} =  \frac{\text{asin}(\mathcal{C})}{2} - \ln\left(\sqrt{1+\sqrt{1-\mathcal{C}^2}}\right) + c = \frac{\text{asin}(\mathcal{C})}{2} - \ln\left(\sqrt{\frac{1+\sqrt{1-\mathcal{C}^2}}{2}}\right),
\end{align}
where in the last line we have chosen the integration constant $c$ such that $\mathcal{C}(0)=0$. We are unaware of an inverse function that would allow us to express $\mathcal{C}$ as a function of $t$ analytically, but the above nevertheless fully specifies the desired solution. By substituting $\mathcal{C}=1$, we determine the hitting time of $t = \pi/4 + \ln(2)/2$.


\subsection{Robustness to loss imbalance and path length fluctuations} 
\label{sec:SMPathLength}

In this section, we calculate the effect of various imperfections in the optical setup. We model loss with additional beam splitters before the which-path-erasing beam splitter, and detector inefficiencies with another pair of beam splitters after it. We use the beam splitter unitary with general transmission and reflection coefficients
\begin{align}
    U_\text{BS} = \begin{pmatrix}
    \sqrt{\eta}_\text{loss}   &   \sqrt{1-\eta_\text{loss}} \\
    \sqrt{1-\eta_\text{loss}}   &   -\sqrt{\eta}_\text{loss}
    \end{pmatrix}.
\end{align}

\begin{figure}
\centering
\includegraphics[width =0.30\textwidth]{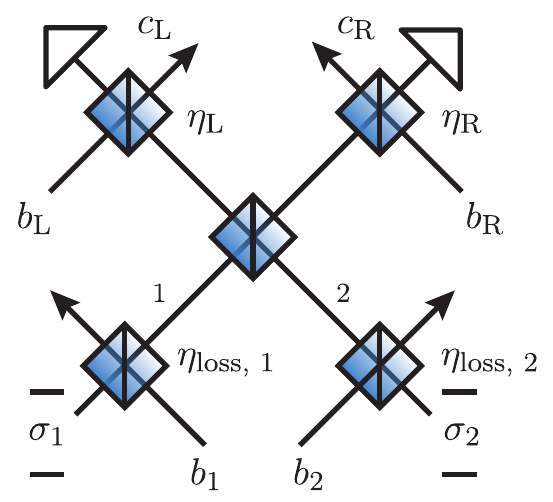}
\caption{A model for the experimental setup of Fig. \ref{fig:PhotonCountingSetup}a, which includes loss, detector inefficiency and path length fluctuations along the connections from the atoms to the beam splitters. All beam splitters used to model loss are fed with optical modes $b_i$ assumed to be in the vacuum state.}
\label{fig:LossImbalance}
\end{figure}

Let us model the effect of each imperfection in turn as it appears in Fig. \ref{fig:LossImbalance}. Path length fluctuations along paths 1 and 2 add overall phase factors to the annihilation operators $\sigma_1$ and $\sigma_2$,
\begin{align}
    \sigma_i &\rightarrow e^{-i\phi_i} \sigma_i, \\ \nonumber
\end{align}
where $\phi_1$ and $\phi_2$ are the optical phases accumulated by the spontaneously emitted photons along each path. Loss along path $i$ may be modelled as mixing in an additional mode $b_i$, assumed to be in the vacuum state
\begin{align}
    \sigma_i &\rightarrow \sigma_i \sqrt{\eta_{\text{loss},i}} + \sqrt{1-\eta_{\text{loss},i}} b_i |0\> = \sigma_i \sqrt{\eta_{\text{loss},i}}.
\end{align}
The signals then reach the physical beam splitter, which acts as
\begin{align}
    c_\text{L/R} = \frac{\sigma_1 \pm \sigma_2}{\sqrt{2}}
\end{align}
as before. After this beam splitter, photon counter efficiencies and subsequent loss may again be modeled with a pair of beam splitters,
\begin{align}
    c_\text{L/R} \rightarrow c_\text{L/R} \sqrt{\eta_\text{L/R}} + \sqrt{1-\eta_\text{L/R}} b_\text{L/R}|0\> = c_\text{L/R} \sqrt{\eta_\text{L/R}}.
\end{align}
Composing all of these transformations, the final jump operators are
\begin{align}
c_\text{L/R} = \sqrt{\eta_\text{L/R}}\frac{\sqrt{\eta_{\text{loss},1}} e^{-i\phi_1} \sigma_1 \pm \sqrt{\eta_{\text{loss}, 2}} e^{-i \phi_2} \sigma_2}{\sqrt{2}}.
\end{align}
Importantly, two applications of $c_\text{L/R}$ to $|ee\>$ with a $\pi$ pulse in between yield the target state $|{\downarrow \uparrow}\> \pm |{\downarrow \uparrow}\>$, since all imbalance parameters fall out as an overall multiplicative factor, resulting in
\begin{align}
    c_\text{L/R} \left(\left( |{\uparrow}\>\<{\downarrow}| + |{\downarrow}\>\<{\uparrow}|\right)^{\otimes 2} \right) c_\text{L/R} |ee\> \propto |{\downarrow \uparrow}\> \pm |{\downarrow \uparrow}\>.
\end{align}

\subsection{Finite $\kappa$ effects}
\label{sec:SMFiniteKappa}

Placing an atom in a cavity can enhance its spontaneous emission rate and increase the fluorescence collection efficiency. However the atom-cavity dynamics introduce a new time scale, which we must account for to quantitatively predict the fidelity of entanglement generation. In this section, we derive an expression for the atomic state after a successful round of our photon counting entanglement scheme. The results show that infidelity due to atom-cavity interactions is negligible in the experimentally relevant regime of wide cavity bandwidth $\kappa$ relative to the atom-cavity interaction strength $g$.

In the rotating wave approximation, the full Hamiltonian including cavity dissipation and atom-cavity interactions is
\begin{align}
    H = \sum_i \frac{g_i}{2} (|{\downarrow}\>\<e|_i a_i^\t + h.c.) - i \frac{\kappa_i}{2} a_i^\t a_i
\end{align}
where $i$ indexes the atoms and their respective cavities. For simplicity we take $g_1 = g_2$ and $\kappa_1=\kappa_2$. $H$ governs the conditional dynamics absent the detection of a photon, so it does not preserve the norm of the state. The operators associated with photon detection now act on the cavities instead of the atoms
\begin{align}
    c_{\text{L}/\text{R}} = \frac{a_1 \pm a_2}{\sqrt{2}}.
\end{align}
Before photon detection, the system evolves as a separable product of two atom-cavity states, so we treat these subsystems separately. We parameterize each subsystem as
\begin{align}
    |\psi_i(t)\> = A_i(t) |e,0\> + B_i(t)|{\downarrow},1\>,
\end{align}
with 0 and 1 counting the cavity photon number. In this subspace, $H$ acts as 
\begin{align}
    |\dot{\psi}_i(t)\> = -i H_i |\psi_i(t)\> = \begin{pmatrix}
    \dot{A}_i(t) \\
    \dot{B}_i(t)
    \end{pmatrix} = \frac{1}{2}\begin{pmatrix}
    0       & -i g \\
    -i g  & -\kappa
    \end{pmatrix},
\end{align}
where $H_i$ only contains terms from the $i$th subsystem. Taking the initial conditions specified by our protocol ($A_i(0) = 1, B_i(0) = 0$), the exact solution is
\begin{align}
    A_i(t) &= \frac{\kappa}{4\sqrt{\kappa^2/4 - g^2}} \left( e^{-\Gamma_\text{slow} t/2} - e^{-\Gamma_\text{fast} t/2}\right) + \frac{1}{2} \left(e^{-\Gamma_\text{slow} t/2} + e^{-\Gamma_\text{fast} t/2} \right) \\ \nonumber
    B_i(t) &= \frac{i g}{2\sqrt{\kappa^2/4 - g^2}} \left( e^{-\Gamma_\text{fast}t/2} - e^{-\Gamma_\text{slow} t/2} \right) \\ \nonumber
    \Gamma_\text{fast} &= \frac{\kappa}{2} + \sqrt{\kappa^2/4 - g^2} ~~~~~ \Gamma_\text{slow} = \frac{\kappa}{2} - \sqrt{\kappa^2/4 - g^2}.
\end{align}
In the experimentally relevant limit of $\kappa \gg g$, this simplifies to
\begin{align}
    A(t) &\approx e^{-\Gamma_\text{slow} t/2} \\ \nonumber
    B(t) &\approx \frac{i g}{2\kappa} \left( e^{-\Gamma_\text{fast}t/2} - e^{-\Gamma_\text{slow} t/2} \right) \\ \nonumber
    \Gamma_\text{fast} &\approx \kappa ~~~~~ \Gamma_\text{slow} \approx \frac{g^2}{\kappa}.
\end{align}
The above simplification makes it clear that $\Gamma_\text{slow}$ may be interpreted as the effective spontaneous emission rate $\gamma$; after an initial transient lasting of order $1/\kappa$, the norm of the state decays exponentially at a rate of $\Gamma_\text{slow}$, indicating increasing probability for detection of a photon.

The full system state immediately before the first photon detection is
\begin{align}
    |\psi(\tau_1 - \epsilon)\> = (A(\tau_1) |e,0\> + B(\tau_1) |{\downarrow},1\>)^{\otimes 2},
\end{align}
where $\epsilon$ is a negligibly small time interval. Immediately after detection, the state becomes
\begin{align}
    |\psi(\tau_1)\> &= c_{\text{L}/\text{R}} |\psi(\tau_1-\epsilon)\> \\ \nonumber
    &= B(\tau_1) |{\downarrow},0\> \otimes (A(\tau_1) |e,0\> + B(\tau_1) |{\downarrow},1\>) \pm (A(\tau_1) |e,0\> + B(\tau_1) |{\downarrow},1\>) \otimes B(\tau_1) |{\downarrow}, 0\>,
\end{align}
where we have dropped a global factor of $1/\sqrt{2}$. The state then continues to evolve under $H$ until feedback exchanges $|{\uparrow}\>$ and $|{\downarrow}\>$. The terms in parenthesis were unmodified by the photon detection event, and so we can continue use the original solution to propagate from $\tau_1$ to $\tau_\text{pulse} = \tau_1 + \delta t$. After the feedback pulse, we have
\begin{align}
    |\psi(\tau_\text{pulse})\> &= |{\uparrow},0\> \otimes (A(\tau_\text{pulse}) |e,0\> + B(\tau_\text{pulse}) |{\uparrow},1\>) \pm (A(\tau_\text{pulse}) |e,0\> + B(\tau_\text{pulse}) |{\uparrow},1\>) \otimes |{\uparrow}, 0\>, \end{align}
where we have dropped a global factor of $B(\tau_1)$. After feedback, the evolution of the $|{\uparrow},1\>$ component decouples from the rest of the wave function,
\begin{align}
    H |{\uparrow}, 1\> = -i \frac{\kappa}{2} |{\uparrow},1\>,
\end{align}
and decays exponentially as $e^{-\kappa t/2}|{\uparrow},1\>$. This fast decay is beneficial, as it decreases overlap with the unentangled state $|{\uparrow \uparrow}\>$. At the same time, the $|e,0\>$ terms evolve according to our original solution. Immediately before the second photon detection event, the state is
\begin{align}
    |\psi(\tau_2 - \epsilon)\> =& |{\uparrow},0\> \otimes \left(A(\tau_\text{pulse}) \left(A(\tau_2-\tau_\text{pulse}) |e,0\> + B(\tau_2 - \tau_\text{pulse}) |{\downarrow},1\>\right) + e^{-\kappa(\tau_2-\tau_\text{pulse})/2} B(\tau_\text{pulse}) |{\uparrow},1\>\right) \\ \nonumber
    &\pm \left(A(\tau_\text{pulse}) \left(A(\tau_2-\tau_\text{pulse}) |e,0\> + B(\tau_2 - \tau_\text{pulse}) |{\downarrow},1\>\right) + e^{-\kappa(\tau_2 - \tau_\text{pulse})/2} B(\tau_\text{pulse}) |{\uparrow},1\>\right) \otimes |{\uparrow}, 0\>.
\end{align}
The final photon detection at $\tau_2$ simply eliminates the $|e,0\>$ component and projects the cavity state into $|00\>$. Factoring out the cavity, the qubits are left in
the state
\begin{align}
    |\psi(\tau_2)\> = A(\tau_\text{pulse}) B(\tau_2 - \tau_\text{pulse}) (\pm_1 |{\downarrow \uparrow}\> \pm_2 |{\uparrow \downarrow}\>) + B(\tau_\text{pulse})e^{-\kappa(\tau_2-\tau_\text{pulse})/2}(\pm_1 1 \pm_2 1) |{\uparrow \uparrow}\>,
\end{align}
where $\pm_i$ indicates the phase associated with the $i$th detection event ($+$ for the left detector and $-$ for the right). If opposite detectors click, then we are left exactly in the singlet state $({\downarrow \uparrow}\> - |{\uparrow \downarrow}\>)/\sqrt{2}$. If the same detector clicks twice, then the exact solutions for $A$ and $B$ derived above suffice to provide an exact expression for $|\psi(\tau_2)\>$. To obtain a simpler result, one can expand the exact solution to first order in $g$, which turns out to be independent of $g$
\begin{align} \label{eq:PsiTau2Approx1}
    |\psi(\tau_2)\> \approx \frac{|{\uparrow \downarrow}\> + |{\downarrow\uparrow}\>}{\sqrt{2}} + \sqrt{2} \frac{1-e^{-\kappa \tau_\text{pulse}/2}}{e^{\kappa(\tau_2 - \tau_\text{pulse})/2}-1} |{\uparrow \uparrow}\>.
\end{align}
Taking the decay rate to be slow compared to $\kappa$, so that we can assume $\tau_\text{pulse},~ \tau_2-\tau_\text{pulse} \gg 1/\kappa$, we arrive at the expression given in the main text,
\begin{align}
    |\psi(\tau_2)\> \approx \frac{|{\uparrow \downarrow}\> + |{\downarrow\uparrow}\>}{\sqrt{2}} + \sqrt{2}e^{-\kappa(\tau_2-\tau_\text{pulse})/2} |{\uparrow \uparrow}\>.
\end{align}
If a detection event happens to occur immediately after preparing $|ee\>$ or immediately after $\tau_\text{pulse}$, then \erf{eq:PsiTau2Approx1} will be more accurate, and there may be significantly more overlap with the unentangled state $|{\uparrow \uparrow}\>$. Fortunately, the probability for this occurrence is suppressed, scaling as $\gamma/\kappa$.

One might hope that the coherence between the Bell state and the undesired $|{\uparrow \uparrow}\>$ component of the wave function could enable an even larger Bell state fidelity using local operations. However, the lack of a $|{\downarrow \downarrow}\>$ component in $|\psi\>$ implies that the concurrence is only a function of the overlap with $|{\uparrow \downarrow}\> + |{\downarrow \uparrow}\>$, as is evident from \erf{eq:concurrence}. Therefore, \erf{eq:PsiTau2Approx1} gives the optimal representation in terms of Bell state fidelity.

\subsection{Locally optimal feedback in the presence of loss and delay}
\label{sec:SMPaQS}

In this section, we derive the locally optimal control operation, which maximizes the fidelity of the current state with respect to a general target state $|\psi_T\>$, or more generally the expectation value of an operator $X_T$. We begin by following the proportional and quantum state (PaQS) feedback scheme of Ref. \cite{Zhang2018Locally}, but going beyond this to include multiple potentially non-commuting feedback Hamiltonians. We further add a correction that accounts for feedback delay \cite{wiseman1994feedback}, and then compute a correction to the locally optimal protocol at first order in the delay. 

\subsubsection{A general expression for the locally optimal measurement-based feedback Hamiltonians}

Let $dW_i$ label the noise corresponding to measurement of operator $M_i$, and $\{H_j\}$ be a set of feedback Hamiltonians. As noted in the main text, we can treat a completely general measurement-based feedback problem by allowing $H_j$ to be a complete matrix basis for the system, so that the locally optimal Hamiltonian is singled out among all possible Hamiltonians. One can also impose experimental constraints like locality by correspondingly restricting $\{H_j\}$ to a subalgebra. The most general form of proportional feedback applies the $j$th feedback Hamiltonian in response to all measurement outcomes using independent proportionality coefficients $A_{ij}$. In the absence of delay, the feedback unitary applied at time $t$ may be written in the form
\begin{align} \label{eq:UNoDelay}
	U(t) &= U(\{\theta_j(t)\}) = \exp \left(-i \sum_j \theta_j(t) H_j\right)
    = I - i\sum_{ij} \tilde{H}_i(A) dW_i(t) - \left[ i\sum_j B_j(t) H_j + \frac{1}{2} \sum_{i} \tilde{H}_i(A)^2 \right]dt \\ \nonumber
    \theta_j(t) &= B_j(t) dt + \sum_i A_{ij}(t) dW_i(t) \\ \nonumber
    \tilde{H}_i(A) &\equiv \sum_j A_{ij} H_j.
\end{align}
We can derive the feedback master equation by applying $U(t)$ to the post-measurement state
\begin{align}   \llabel{eq:drhoM}
    \rho(t+dt) &= U(t) \rho_M(t) U^\dagger(t) \\ \nonumber
    \rho_M(t) &\equiv \rho(t) + \sum_i \mathcal{D}[M_i]\rho(t) dt + \sqrt{\eta_i}\mathcal{H}[M_i]\rho(t) dW_i(t).
\end{align}
Using the expansion \erf{eq:UNoDelay} and applying Ito's rule immediately gives
\begin{align}   \llabel{eq:rhoMNoDelay}
    d\rho(t) &= - i\sum_j B_j[H_j,\rho(t)]dt \\ \nonumber
    & ~~~ + \sum_i \Big[ \mathcal{D}[M_i]\rho(t)dt + \sqrt{\eta_i}\mathcal{H}[M_i]\rho(t)dW_i(t) \\ \nonumber
    & ~~~ - i [\tilde{H}_i(A), \rho(t)]dW_i(t) + \mathcal{D}[\tilde{H}_i(A)]\rho(t)dt - i\sqrt{\eta_i} [\tilde{H}_i(A), \mathcal{H}[M_i]\rho(t)] dt \Big]  \nonumber \\
    \rho(t+dt) &= \rho(t) + d\rho(t).
\end{align}
We now solve for the locally optimal protocol, using the expectation value $\Tr[X_T \rho(t + dt)]$ as our figure of merit. Simply maximizing $\Tr[X_T d\rho]$ over $A_{ij}$ and $B_j$ does not work; $d\rho$ is linear in $B_j$, so the maximization condition $\partial \Tr[X_T \rho(t+dt)]/\partial B_j = -i \Tr[X_T[H_j, \rho(t)]]dt = 0$ does not depend on $B_j$ and is therefore typically inconsistent. Instead, we must first allow for arbitrarily large feedback rotations $\theta_j$ to compute a general condition for maximization, and only then assume that $\theta_j$ is infinitesimal.

Returning to the first line of \erf{eq:rhoMNoDelay}, the derivative of $\rho(t+dt)$ with respect to $\theta_a$, which we use to maximize $\Tr[X_T \rho(t+dt)]$, is
\begin{align}
    \frac{\partial \rho(t+dt)}{\partial \theta_a}  &= \frac{\partial}{\partial \theta_a} \left(U \rho_M U^\dagger\right) = Q_a \rho(t+dt) + \rho(t+dt) Q_a^\dagger \\ \nonumber
    Q_a &\equiv \left(\frac{\partial}{\partial \theta_a} U\right) U^\dagger \\ \nonumber
    &= -i H_a + \frac{1}{2} \sum_i \theta_i [H_i, H_a] + \frac{i}{6} \sum_{ij}\theta_i \theta_j [H_i, [H_j, H_a]] + ...
\end{align}
The expression for $Q_a$ may be derived using the useful identity $d e^{X(t)}/dt = -\int_0^1 e^{sX(t)}(dX(t)/dt)e^{(1-s)X(t)}ds$ and the Baker-Campbell-Hausdorff formula\cite{lutzky1968parameter}. We can then write the maximization condition of $\Tr[\rho(t+dt) X_T]$ with respect to $\theta_a$ as
\begin{align}   \llabel{eq:dTrdTheta}
    \frac{\6 \Tr[X_T \rho(t+dt)]}{\6 \theta_a} = &-i \<[H_a, \rho(t)]\>_{X_T} + \frac{1}{2}\sum_j \theta_j \<[[H_j, H_a], \rho(t)]\>_{X_T} + \frac{i}{6} \sum_{jk} \theta_j \theta_k \<[[H_j, [H_k, H_a]],\rho(t)]\>_{X_T} \\ \nonumber
    &-i \<[H_a, d\rho]\>_{X_T} + \frac{1}{2}\sum_j \theta_j \<[[H_j, H_a], d\rho]\>_{X_T} = 0,
\end{align}
where we use the shorthand $\<\mathcal{O}\>_{X_T} \equiv \Tr[\mathcal{O} X_T]$ and have dropped all terms of order $dt dW_i$ and higher. In general, \erf{eq:dTrdTheta} will not admit an analytic solution. In fact, if the first term is non-zero, the maximization condition becomes inconsistent, because all other terms are infinitesimal by assumption.

To derive the feedback coefficients from \erf{eq:dTrdTheta}, we must assume that an (at least locally) optimal rotation was applied at the previous time step $t-dt$. This could require a finite feedback rotation, particularly at the first time step of a feedback protocol when starting with a generic state. In general, requiring 
\begin{align}   \llabel{eq:GeneralOptCondition}
    \left\< \frac{\6 U(\{\theta_j(t-dt)\}) \rho_M(t-dt) U(\{\theta_j(t-dt)\})^\t}{\6 \theta_a}\right\>_{X_T} = \left< \frac{\6 \rho(t)}{\6 \theta_a} \right\>_{X_T} = 0
\end{align}
is not a sufficient condition for having found a local optimum. If the set of all $U$s generated by \erf{eq:UNoDelay} allowing for all values of $\theta_j$ do not form a group, then it is possible to satisfy \erf{eq:GeneralOptCondition} but not to have found a local maximum \textit{i.e.}, 
\begin{align} \llabel{eq:SecondOpt}
    \left. \left\< \frac{\6 U(\{\theta'_j\})\rho(t)U(\{\theta'_j\})^\t}{\6 \theta'_b} \right\>_{X_T} \right|_{\theta'_j=0} \neq 0
\end{align}
%
for some subsequent rotation $\theta'_j$. Lack of closure of $U$ under multiplication allows $U(\{\theta'_j\}) U(\{\theta_j\})$ to yield new operations that cannot be written as $U(\{\theta_j\})$. We can enforce that $U$ forms a group by requiring that $\{H_j\}$ forms a Lie algebra. This mandates that there exist structure constants $f_{ij}^k$ such that $[H_i, H_j] = \sum_{k} f_{ij}^k H_k$ for all $i$ and $j$. This constraint is not severe in practice, at least for finite Hilbert spaces; given a set of Hamiltonians that do not form a Lie algebra, one can construct the Lie algebraic closure by augmenting this set with commutators of the Hamiltonians already in the set until we can no longer generate linearly independent operators.

Given that $\{H_j\}$ forms a Lie algebra, the assumption that the locally optimal rotation was applied at the previous time step implies that \erf{eq:SecondOpt} goes to zero. This immediately implies that the first term of \erf{eq:dTrdTheta} drops out. Furthermore, as $[H_i,H_a]$ and $[H_i, [H_j, H_a]]$ can be written as linear combinations of $H_j$, the second and third terms drop out as well, leaving us with
\begin{align} \llabel{eq:dTrdTheta2}
    \frac{\6 \Tr[X_T \rho(t+dt)]}{\6 \theta_a} = -i \<[H_a, d\rho]\>_{X_T} + \frac{1}{2}\sum_i \<[[\tilde{H}_i, H_a], d\rho]\>_{X_T} dW_i = 0,
\end{align}
where we have expanded $\theta_j$ and dropped higher-order terms. We can now solve for $A_{ij}$ and $B_j$ order-by-order in $dW_i$. First considering terms of order $dW_i$, we find
\begin{align} \llabel{eq:PaQSdWTerms}
     \underbrace{\sqrt{\eta_i} \<[H_a, \mathcal{H}[M_i]\rho(t)]\>_{X_T}}_{a_{ia}} -i \sum_j &A_{ij} \underbrace{\<[H_a, [H_j, \rho(t)]]\>_{X_T}}_{-c_{ja}} = 0, \\ \nonumber
    A &= i a c^{-1},
\end{align}
in agreement with the main text. For the $dt$ terms, note that the second term of \erf{eq:dTrdTheta2} picks out $\mathcal{O}(dW_i)$ terms of $d\rho$. \erf{eq:PaQSdWTerms} implies that this term is already zero. Thus the $\mathcal{O}(dt)$ terms of \erf{eq:dTrdTheta2} give
\begin{align}   \llabel{eq:PaQSdtTerms}
    -i \sum_j &B_j \underbrace{\<[H_a, [H_j, \rho(t)]] \>_{X_T}}_{-c_{ja}} + \underbrace{\sum_i \< [H_a, \mathcal{D}[M_i]\rho(t) + \mathcal{D}[\tilde{H_i}]\rho(t) - i\sqrt{\eta_i} [\tilde{H}_i, \mathcal{H}[M_i] \rho(t)] ] \>_{X_T}}_{b_a} = 0, \\ \nonumber
    B &= i b c^{-1},
\end{align}
treating $b$ as a row vector.

Together, \erf{eq:PaQSdWTerms} and \erf{eq:PaQSdtTerms} form an exact, closed form expression for the feedback controls that extremize the cost function under measurement-based feedback, as modelled by \erf{eq:rhoMNoDelay}. The condition to have found a local maximum as opposed to a minimum or saddle point is that the Hessian matrix
\begin{align} \label{eq:PaQSHessian}
    \frac{\partial^2}{\6 \theta_a \6\theta_b} \<\rho(t+dt)\>_{X_T}
\end{align}
should be negative definite. We can compute it exactly as
\begin{align}
    \frac{\partial^2 \rho(t+dt)}{\6\theta_a \6\theta_b} &= Q_{a b}\rho(t+dt) + \rho(t+dt) Q^\dagger_{a b}  + Q_a \rho(t+dt) Q_b^\dagger + Q_b \rho(t+dt) Q_a^\dagger \\ \nonumber
   Q_{a b} &\equiv \frac{\6^2 U}{\6\theta_a \6 \theta_b} U^\t = \frac{\6(Q_b U)}{\6\theta_a}U^\t = \frac{\6 Q_b}{\6\theta_a} + Q_b Q_a.
\end{align}
In principle, one could use the above formula to compute the Hessian exactly in terms of $\rho(t+dt)$, $M_i$ and $H_j$. However unlike in the first derivative, there are typically both finite and infinitesimal contributions here, so only the finite part is essential. Taking the derivative of $Q_b$ and then dropping terms of order $dW$ and $dt$ yields
\begin{align}
    \frac{\6^2}{\6 \theta_a \6 \theta_b} \<\rho(t+dt)\>_{X_T} = -\< ([H_b, [H_a, \rho(t+dt)]])\>_{X_T} + \mathcal{O}(dW) \approx c_{ab},
\end{align}
where we have eliminated a term of the form $\<[[H_a, H_b], \rho(t+dt)]\>_{X_T}$ and $c_{ab}$ was defined in \erf{eq:PaQSdWTerms}.

If $c$ fails to be negative definite, then it is often necessary to perform a global search over $\theta_j$ to find the global maximum of the figure of merit. An important exception to this requirement is when a linear combination of the $H_i$s commutes with $\rho$, in which case $c$ will have a null eigenvalue. In this case, a global search may not improve the figure of merit, as $\rho$ is invariant under this control operation. Thus it is preferable to simply ignore null eigenvalues of $c$ and only insist that it is negative semidefinite. To do so, one may simply replace the matrix inverse in \erf{eq:PaQSdWTerms} and \erf{eq:PaQSdtTerms} with a Moore-Penrose pseudoinverse \cite{campbell2009generalized}, which is what we use in practice.
In general, the above expression guarantees that we have found a local maximum of the figure of merit, but this may not coincide with a global maximum. To check the latter, one must occasionally perform global searches over $\theta_j$ allowing for large rotation angles. These global searches are often not necessary to achieve good results, and were not implemented in the protocol presented here.


\subsubsection{Locally optimal feedback in the presence of feedback delay}

As a final step, we now derive corrections to the master equation accounting for feedback delay along the lines of Ref. \cite{wiseman1994feedback}, and then derive corrections to $A$ and $B$ which retain local optimality to first order in the delay parameter $\delta t$. Let us repeat the derivation of the feedback master equation, but enforce $\theta(t)$ to depend on the measurement noise at $t-\delta t$
\begin{align} \llabel{eq:ThetaDelay}
    \theta_j(t) &= B_j(t) dt + \sum_i A_{ij}(t) dW_i(t-\delta t).
\end{align}
Measurement at the current time does not correlate with feedback at the current time, meaning that terms of the form $dW_i(t) dW_j(t-\delta t)$ drop. Expanding \erf{eq:drhoM} and \erf{eq:UNoDelay} using \erf{eq:ThetaDelay} gives
\begin{align}   \llabel{eq:SMEDelay}
    \rho(t+dt) = &\rho(t) - i\sum_j B_j[H_j,\rho(t)]dt + \sum_i \Big[ \mathcal{D}[M_i]\rho(t) dt + \sqrt{\eta_i}\mathcal{H}[M_i]\rho(t) dW_i(t)  - i [\tilde{H}_i,\rho(t)]dW_i(t-\delta t) 
    + \mathcal{D}[\tilde{H}_i]\rho(t) \Big].
\end{align}
To capture the effects of delay, we need to explicitly write down all $dW_i(t-\delta t)$ terms so that we can compute the resulting cross terms.
Feedback at time $t$ correlates with the measurement noise at time $t-\delta t$, so we expand $\rho(t)$ to capture its $dW_i(t-\delta t)$ dependence and then substitute the result into the feedback term $[H_j,\rho(t)]dW_i(t-\delta t)$. Expanding $\rho(t)$ yields
\begin{align} \llabel{eq:RhotSubDerive}
    \rho(t) &= e^{(\delta t-dt)\mathcal{L}}\Big[ I + \sum_i\sqrt{\eta_i}\mathcal{H}[M_i]dW_i(t-\delta t) + ...\Big]\rho(t-\delta t) \\ \nonumber 
    &= \rho(t-dt) + \sum_i e^{(\delta t -dt)\mathcal{L}}  \sqrt{\eta_i}\mathcal{H}[M_i]dW_i(t-\delta t) \rho(t-\delta t)+ ... \\ \nonumber
    &= \rho(t) + \sum_i e^{\delta t \mathcal{L}}\sqrt{\eta_i}\mathcal{H}[M_i]dW_i(t-\delta t)\rho(t-\delta t) + \mathcal{O}(dt, dW_i(t-dt), dW_i(t-\delta t - dt)) + ...,
    \end{align}
where $e^{\delta t \mathcal{L}}$ is the time evolution operator from $t-\delta t$ to $t$ (which we approximate to be linear)
and we have only explicitly written terms that will not drop in the end upon application of the Ito rules. The $\mathcal{O}$ terms come from the fact that in the last line, we have propagated one time step further, from $t-dt$ to $t$, unlike in the previous line. The difference cancels on substitution and allows us to write everything in terms of $\rho(t)$ and $\rho(t-\delta t)$. We can simplify somewhat further if we take $\delta t$ to be small compared to the timescale of the measurement and feedback dynamics. Using $\rho(t-\delta t) = e^{-\delta t\mathcal{L}} \rho(t)$ and $e^{\pm\delta t \mathcal{L}} \approx (1 \pm \delta t \mathcal{L})$, we can approximate the above expression as
\begin{align} \llabel{eq:RhotSub}
    \rho(t) \approx \rho(t) + \sum_i \sqrt{\eta_i} \left[ \mathcal{H}[M_i]\rho(t) + \delta t(\mathcal{L}\mathcal{H}[M_i]\rho(t) - \mathcal{H}[M_i]\mathcal{L}\rho(t)) \right] dW_i(t-\delta t).
\end{align}
Substituting the above $\rho(t)$ into the aformentioned feedback term in which correlations arise yields
\begin{align} 
   \sum_{i} & [\tilde{H}_i, \rho(t)]dW_i(t-\delta t) \\ \nonumber 
%
%
%
    &\approx \sum_{i} [\tilde{H}_i, \rho(t)] dW_i(t-\delta t) + \sqrt{\eta_i}[\tilde{H}_i, \bar{\mathcal{H}}[M_i]\rho(t)] dt + \delta t\sqrt{\eta_i} [\tilde{H}_i, \mathcal{L}\bar{\mathcal{H}}[M_i]\rho(t) - \bar{\mathcal{H}}[M_i]\mathcal{L}\rho(t)] dt,
\end{align}
where we have defined $\bar{\mathcal{H}}[X]\rho = X \rho + \rho X^\t$. The intuitive interpretation of the above term, evident from \erf{eq:RhotSubDerive} and \erf{eq:RhotSub} is that we have taken the state, evolved backward in time, applied measurement, evolved forward in time, and then applied the corresponding feedback operation. 

Putting everything together, we arrive at a final master equation
\begin{align}
     \rho(t+dt) = &\rho(t) -i\sum_j B_j[H_j,\rho(t)]dt + \sum_i \Big[ \mathcal{D}[M_i]\rho(t) dt + \sqrt{\eta_i}\mathcal{H}[M_i]\rho(t) dW_i(t)  \\ \nonumber
    &-i \Big( [\tilde{H}_i, \rho(t)] dW_i(t-\delta t) + \sqrt{\eta_i}[\tilde{H}_i, \bar{\mathcal{H}}[M_i]\rho(t)] dt + \delta t\sqrt{\eta_i} [\tilde{H}_i, \mathcal{L}\bar{\mathcal{H}}[M_i]\rho(t) - \bar{\mathcal{H}}[M_i]\mathcal{L}\rho(t)] dt \Big) \\ \nonumber
    &+ \mathcal{D}[\tilde{H}_i]\rho(t) \Big],
\end{align}
which is approximately Markovian. 
In particular, since we have captured all of the noise correlations explicitly, it is valid to average over measurement outcomes by dropping $dW$ terms, as in a typical master equation.

To compute the optimization condition, we use \erf{eq:dTrdTheta} to compute the derivatives with respect to $\theta_a$, though this time the $dW_i$ component of $\theta_a(t)$ corresponds to the earlier time $t-\delta t$. We assume that $A$ and $B$ are relatively constant over the delay interval $\delta t$ \textit{i.e.}, $A(t) \approx A(t-\delta t)$ and $B(t) \approx B(t-\delta t)$. As before, we assume that the optimal rotation was applied at the previous time step, but this time averaging over all measurement outcomes that have not yet been observed due to delay (from $t-\delta t + dt$ to $t$)
\begin{align}   \llabel{eq:DelayPaQSMaxCondition}
    E&_{dW_i(t-\delta t + dt)... dW_i(t)}\left[\frac{\6 \<\rho(t+dt)\>_{X_T}}{\6 \theta_a}\right] = 0. \\ \nonumber
\end{align}
We now solve the maximization condition order-by-order. $dW_i(t)$ terms average out, as there are no feedback terms depending on $dW_i(t)$. The $\mathcal{O}(1)$ term drops along with nested commutators containing $\rho(t)$ (second and third terms of \erf{eq:dTrdTheta}), as we assume that the locally optimal rotation was applied at the previous time step. For the $\mathcal{O}(dW_i(t-\delta t))$ terms, we must pull any $dW_i(t-\delta t)$ dependence out of $\rho(t)$ in \erf{eq:SMEDelay}. Using \erf{eq:RhotSub}, the maximization condition is
\begin{align}
    \underbrace{\sqrt{\eta_i} \<[H_a, \bar{\mathcal{H}}[M_i]\rho(t)]\>_{X_T}}_{a_{ia}} + \delta t\underbrace{ \sqrt{\eta_i} \<[H_a, \mathcal{L}\bar{\mathcal{H}}[M_i]\rho(t) - \bar{\mathcal{H}}[M_i]\mathcal{L}\rho(t)]\>_{X_T}}_{a'_{ia}} - i \sum_j A_{ij} \underbrace{\<[H_a, [H_j, \rho(t)]]\>_{X_T}}_{-c_{ja}} = 0
\end{align}
To zeroth-order in $\delta t$, \erf{eq:PaQSdWTerms} solves the above equation. The first-order correction is
\begin{align}   \llabel{eq:AWithDelay}
    A = i(a + \delta t a')c^{-1}
\end{align}
$a'$ depends on both $A$ and $B$ through its dependence on $\mathcal{L}$, but \erf{eq:AWithDelay} is still correct to first order if one uses the zeroth-order solution \erf{eq:PaQSdWTerms} and \erf{eq:PaQSdtTerms} in $\mathcal{L}$. Finally, the $\mathcal{O}(dt)$ terms of \erf{eq:DelayPaQSMaxCondition} gives a solution for $B$ in the same way
\begin{align}
    -i \sum_j \< [H_a, [H_j, \rho(t)]]\>_{X_T} &+ \underbrace{\<[H_a, \sum_i \mathcal{D}[M_i]\rho(t) - i \sqrt{\eta_i}[\tilde{H}_i, \bar{\mathcal{H}}[M_i]\rho(t)] + \mathcal{D}[\tilde{H}_i]\rho(t)]\>_{X_T}}_{b_a} \\ \nonumber 
    &\underbrace{- i \< [H_a, \sum_i \sqrt{\eta_i} [\tilde{H}_i, \mathcal{L}\bar{\mathcal{H}}[M_i]\rho(t) - \bar{\mathcal{H}}[M_i]\mathcal{L}\rho(t)]] \>_{X_T}}_{b'_a} \delta t \\ \nonumber
    B &= -i (b + \delta t b')c^{-1}.
\end{align}
These relations give the locally optimal feedback protocol for a general measurement-based scheme in the presence of delay.

\clearpage

\end{document}